# Consciousness and Learning based on DNA Recombination and Memristor Quality of Microtubules.

Axel Dietrich[1*]  &  Willem Been[†]

)[1] Retired from the Institute of Human Genetics of the University of Amsterdam, the Netherlands.
)[†] Willem Been died in 2011 still working at the Department of Anatomy and Embryology of the University of Amsterdam. He is co-author to honour him and because of his considerable contribution to this paper

)[*] Corresponding address:  axdietrich@gmail.com

**Abstract**

This paper is a completion of an earlier model proposed by us. In the model different memories are attached at cell surface determinants which are the result of DNA recombination.
Our earlier experiments strongly suggest that DNA recombination actually takes place during a short period of early development in the brain in a limited number of neurons. In the present paper a model is presented in which switchboard neurons play a key role in the storage and retrieving of memory. And as a consequence, they play a major role in the process of learning and form the basic material for consciousness.
In the original model there was insufficient explanation for the realization of the internal connection of one cell surface determinant to the other. We realized that tubulin should play a role in these intracellular connections. The tubulin molecules can form a connective wire because of a change of shape of the individual tubulin dimers. This way the fast switch is realized by the switch of the tubulin dimer configuration. Because the cell should remember which switch was activated and which one was not or less activated, we postulate the memristor quality of microtubules.

**Introduction**

What is consciousness? Already for a long time there has been quite a lot of publishing about consciousness, for example: In one single decade there were at least 30,000 publications dealing with consciousness (Carter, 2002). It is in fact not explained what consciousness actually is and the biggest unanswered question is how the brain generates consciousness (Greenfield, 2004; Miller, 2005). Perhaps it is even impossible to define consciousness and it will probably be better to just work with it as put by Michael Gazzaniga in his words "You don't waste your time defining the thing. You just go out and study it" (Ledford, 2008). We will keep this in mind when we make some statements about consciousness, in spite of the poor definition or even the total absence of definition.

The one thing we are convinced of is that the basis for consciousness is memory and the association of different memories in itself forms the basis for learning and it is even possible to use this as the definition of learning, as Eccles (1986) already pointed out: "Memory of some kind is required for all conscious experiences and actions". This brings us to the underlying problem: What is memory?

It is possible to distinguish different types of memories such as short-term memory, long-term memory, declarative (or explicit) memory, nondeclarative (or implicit) memory, procedural memory, priming, conditioning and nonassociative learning etc. etc. (Milner et al.,1998). In most cases description takes care of the definition, but in general memory can be defined as: "A memory is a physical system for transferring information from one moment in time to another, where that information concerns something external to the system itself" (Wolpert, 1992). Safe, but we prefer to use a more general working definition: Memory is any kind of storage of information.

In spite of the fact that some authors (Forsdyke, 2009) discuss the possibility that memory could be stored outside the brain, it is generally accepted that the localization of what we call memory is in the central nerve system, which was clearly demonstrated by the Penfield/Milner experiments (Penfield & Milner, 1958; Milner et al., 1998).
To avoid misunderstanding: the storage of information in other systems than the nervous system such as the immune system and genetical systems is sometimes also defined as memory, but that type of "memory" is stored in locations which are specific for the system itself and will not be discussed here in spite of the fact that it is, in our vision, real memory.
The association of memories is the item of this paper and it covers a large field. Learning is one of the forms of association of experiences. This is demonstrated by Seitz et al. (2009). In their experiments the learning process is clearly demonstrated by unconscious association of different experiences.

Of course there is a paradigm concerning memory storage. In this paradigm it is accepted that neural plasticity and the formation of new neural synapses and the alteration of their strength forms the basis of the storage of memories.

It all started with Santiago Ramón y Cajal, who proposed the neuron model with synapses as connection for the basis of nervous system (Cajal, 1894). He formulated the principles which later gave rise to the neuron doctrine in which the neuron forms the fundamental structural and functional unit of the brain (Kandel, 2006). He also suggested that neurons communicate with other neurons at specialized sites, which later were named synapses. Cajal also pointed out that these communication sites should be highly specific and he suggested that the internal pathways and connections in the brain alter with experience during the process of learning. This model of neural plasticity was later in 1949 proposed by Hebb and is considered, up until now, the basis for memory storage, learning and consciousness (Huang, 2008). Hebb emphasized the changes in synapses that can be observed depending on the frequency and the intensity in which certain connections are used, or not used in which case there will be a decrease of the connection.

Up until now this model of plasticity in synapses is general accepted, but there are at least three arguments against a model which is based purely on the plasticity of neural synaptic networks: 1) complex structures are necessary, 2) plasticity results in instability and 3) retrieval of memory is difficult to explain.

**1)** For the functioning of systems that use synaptic plasticity, complex structures are necessary to facilitate complicated network structures, however there is a large number of animal species with memory and a learning abilities, without complex nervous structures such as the neocortex and a hippocampus.

For thousands of years we know, from the training of our fellow mammals such as dogs and horses, that these animals have a memory and learning system, which strongly resembles ours.

However, also simpler animals, with simple nerve systems, demonstrate the storage of memories and the ability of a learning process. And from this it must be concluded that it will be necessary to perform it without a complex network and that single neurons probably play a role in the process of memory storage.

For example, insects also possess a relatively simple nerve centre. In one group of insects: the honeybees it was demonstrated that they are able to learn to distinguish different human faces (Dyer et al., 2005).

Karl von Frisch in his famous research, also of the honeybees, for which he was rewarded the Nobel price in 1973, indicated that he always took it for granted that honey bees possess facilities for memory storage and learning (von Frisch, 1965).

Recently (Chabaud et al., 2009) these facilities of memory storage and learning were demonstrated in other insects. Memory storage and learning in other animals than insects, but with a simple nerve system, were also demonstrated for example in the marine snail *Aplysia* (Kupfermann & Kandel, 1969; Kandel, 2001). And again it was concluded that it is inevitable that single neurons play a role in these kind of processes (Arshavsky, 2001; Dietrich & Been, 2001).

Another example is the nautilus, an archaic relative of octopuses, cuttlefishes and squids (coleoid cephalopods). It has the capacity for memory storage and learning. The coleoid cephalopods have specialised complex brains containing dedicated learning and memory centres. The primitive nautilus can be considered as a remnant of an ancient lineage that persisted since the Cambrian. Nautilus brains are relatively simple and the dedicated learning and memory regions are absent. However, despite this lacking of regions that support learning and memory nautilus expressed a similar memory profile as the other coleoids (Crook and Basil, 2008).

All given examples refer to animals, but even among the most primitive forms of life without a nerve centre, systems for memory and learning can be observed, but we will discuss them hereafter.

**2)** Plasticity implies change which will result in instability caused by altering memories. They will change during time. This is undesirable for memories that should be stored in the human brain unaltered (we hope) for one hundred years.

There are actually indications that there must be permanent sectors in the field of memory storage and lifelong memories are associated with stable dendritic spines (Yang et al., 2009; Roberts et al., 2010). In addition, the phenomenon of the recovery of "lost" memories (Fischer et al. 2007) gives a strong indication for a permanent factor in the storing machine. The lost memories must be still there and they did not change after all.

This does not exclude the possibility of plasticity in certain connecting structures such as the hippocampus and the amygdala because these probably do not have a hardware storing function, but can be considered as hatches to and from the actual storing locations.

**3)** As last point it is remarkable that, in most publications on the subject of memory storage, the subject of the retrieval of those memories is ignored, probably because it still is difficult to explain the retrieval of memories especially in the context of neural plasticity.

Summarizing we can state that plasticity of neural connections will be responsible for instability and a model based on such a system makes the explanation of lifelong conservation, storage and retrieval of memories very difficult, which makes this model rather useless. That brought us at a certain moment to the presentation of a model in which single cells and DNA recombination play a role (Dietrich & Been, 2001). Consolidation takes place by specific DNA sequences. These DNA sequences are created by the recombination of DNA in a similar way as during meiosis or the production of immunological antibodies.

DNA has the potential of the production of large numbers of specific DNA sequences and these can function as markers of neural networks images.

There were at that moment a number of considerations that lead to that theory:

-DNA is a very stable molecule and in addition, there are cellular mechanisms operational to maintain the original sequence of the DNA, which guaranties a unchanged sequence.

-Most of the DNA is not used for protein coding sequences: approximately 3% of our DNA are the exon sequences and the rest (97%) is not protein coding probably with other functions or even with no obvious function, in which case we have the undesirable tendency to call it junk DNA. Anyhow, we may conclude that there is sufficient material to play a role in the storage of information.

-There are no cell divisions in the brain after the adulthood is reached. Structural DNA arrangements will not be altered nor disrupted as a consequence of cell division and mitosis.

-Chromosomal pairing was actually demonstrated in the brain, which could be an indication for the exchange of DNA. In brain tissue pairing of homologous chromosomal areas was observed (Arnoldus et al., 1989). This provides a strong indication for genetic recombination analogous with the crossing over which produces meiotic recombination.

In our model the DNA recombination finds its expression in cell surface determinants (neural receptors). These determinants are present in advance and therefore before they are used as an attachment for a memory. This strongly resembles the situation in the immune system. These determinants are supposed to be limited to a special group of neurons that will function in the realization of associations between networks. Because these neurons produce selective switches we call these neurons: switchboard neurons.

**Testing the hypothesis**

In a pilot study of us, using antibodies against a component known to be highly specific for recombination during meiotic stages (Been & Dietrich, 2004) a strong indication for recombination of DNA in the brain was obtained. We observed it in the brain in a number of brain cells, during a discrete period of the early development of mice with the highest frequency in a short period from 5 days after birth until 3 weeks after birth. This is in agreement with the general idea of a critical period in brain development and the coinciding of our data with those from other research is strikingly similar (Hensch, 2004; Roberts et al., 2010). Probably such a critical period is relevant for the handling of memory.

Most likely this is the period in which the cell surface determinants are positioned on the outside of the neuron as a product of DNA recombination (Figure 1 A), just like the V(D)J recombination system which is able, with a limited number of genes using DNA rearrangements, to produce an enormous variety of cell surface determinants.

The gene transcription of the new DNA recombination products can be carried out, using the standard transcription factors that are already present, such as the CREB system.

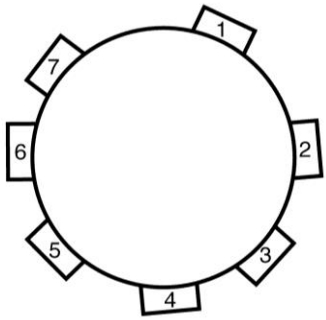

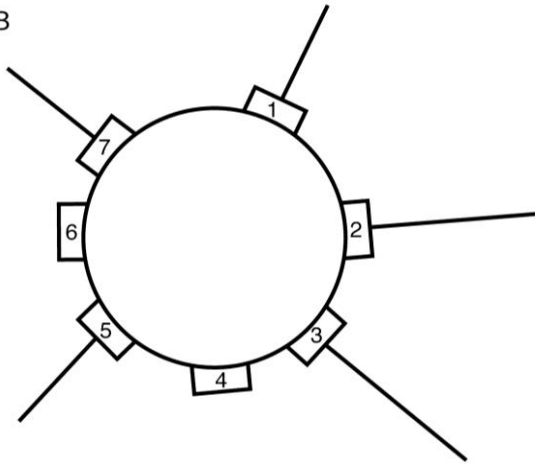

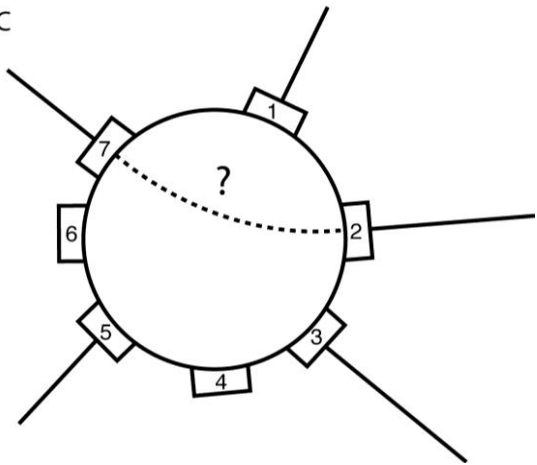

**Figure 1.**

Schematic representation of a switchboard neuron with **A:** The possibility for several connections of neural networks given by the cell surface determinants 1 – 7, which result from DNA recombination. **B:** Different connections are occupied at the positions 1, 2, 3, 5 and 7; the open positions can be used later in life. **C:** It is possible that different networks are recombined, for example 2 with 7.

From that time onwards these cell surface determinants are available for the attachment of different neural networks that function as the carrier of an incoming experience, which can be stored as a memory.

The process of learning extends over the rest of the life of the individual. During life learning is established by the association of different experiences, that are translated into neural networks. These networks can be attached onto the available cell surface determinants of the switchboard neuron (Figure 1B).

In a network all types of information can be stored such as pictures, sound, smell and taste and by means of association they can be linked on another network containing another experience. This is what we call in general memory and sometimes more specific the process of learning. One of the most famous classical examples from literature, which is even discussed in prestigious scientific journals such as Nature (Beauchamp, 2008), is the associations and self-reflection described by Marcel Proust in: "a la recherche du temps perdu" (1919), he describes that when his mother gives him madeleines, sopped in lime-blossom tea, this causes the reliving of considerable sections of his youth, when he also got madeleines sopped in lime-blossom tea from his aunt Léonie. We probably all are familiar with this type of associations.

How is this kind of associations actually made concrete ?

We think that the association between different neural networks is actually made within the switchboard neuron especially because of the results of our immunofluorescent experiment in mouse brain. There we found a strong indication for recombination in the brain using antibodies against components of the recombination nodules which are known to be responsible for meiotic recombination in yeast, plants and animals. We observed a positive reaction in the brain in a limited number of brain cells, during a discrete period of the early development of mice. These results confirmed the earlier presented model of memory storage in which DNA recombination plays a major role. It is also in agreement with the idea that the label for memories is produced in advance and by doing so before the actual experience happens, on a limited number of switchboard neurons.

However, it was not clear at all to us how the switchboard neuron actually carries out the physical connection between two (or more) cell surface determinants on the surface of one cell. Or to put it simple: How is this internal connection performed (Figure 1C)?

What we needed was a connection that shows different activation states and that is capable of "remembering" those states. The answer can be found in the so called memristor principle. What is a memristor? The discussion about memristors started with a paper in 1971 by Chua, who purely theoretical, reasoned that there should exist a fundamental electronic circuit element and he called it the memristor, because it should have properties like resistor property combined with memory qualities.

As we mentioned before there are primitive organisms, without a classical nerve system, that possess a system with brain-like functions such as memory and anticipation. In the slime mold *Physarum* this was convincingly demonstrated (Saigusa et al., 2008). Pershin et al. (2009) found the explanation for these qualities in a memristor structure in the slime mold *Physarum polycephalum*. The ectoplasm of these slime molds contain radial and longitudinal actin-myosin fibers, which have a function in movement, but also in the processing of memory. Pershin and DiVentra (2009) were also able to demonstrate experimentally the formation of associative memory in a simple neural network consisting of three electronic neurons connected by two memristor-emulator synapses.

We think that we can give a satisfactory solution for the shortcoming of our model and we think that tubulin in the configuration of microtubules plays a role. For the following reasons:

**1:** The neurons are replete with microtubules made of tubulin ( Margulis, 1998; Rasmussen et al.,1990). The tubulin is present in practically the entire cell body. It can connect one side of the cell with another side of it. Because of the lack of cell divisions in the adult brain tubulin can fulfill other functions than mitotic spindle and possible connections stay intact.

**2:** Tubulin is associated with memory and learning as demonstrated by Mileusnic et al. (1980). They observed an increase in tubulin in an area of the chick brain following training on passive avoidance learning. Cronly-Dillon et al. (1974) also observed a strong tubulin production in goldfish during a period of learning. It was found in rats, that when the critical learning phase is over, the production of tubulin is drastically reduced (Rasmussen et al., 1990).

**3:** A further indication for the functioning of microtubules in the memory storage is the fact that in Alzheimer's disease there are malfunctioning microtubules in tangles. These are caused by hyperphosphorylation of the Tau protein (Goedert et al., 2006).

Cytoskeletal proteins such as **m**icrotubule **a**ssociated **p**roteins (the MAP/Tau family; Dehmelt and Halpain, 2004) play a role in connecting microtubules to other microtubules. MAP2 and Tau are specifically found in neurons. They have a microtubule stabilizing activity. The phosphorylation and dephosphorylation of MAP2 and Tau regulates their activation.

**4:** Tubulin can have a moving function like the actin-myosin fibers in *Physarum*, but in neurons the moving function is mainly performed by actin. The tubulin will probably have other functions (Tanaka & Sabry, 1995). In addition tubulin can perform a memristor quality like the actin-myosin in *Physarum* because it has the capacity in the shape of a tubule to remember a configuration as described by Mullin (2009).

The mechanism of the actual switch could work as follows:

The components of which microtubules are built of are the tubulin subunits ordered in the microtubule. These subunits are tubulin dimers, which can appear in two conformations; an alpha and a beta state and they can switch from one state to the other.

This switch from one state to another is caused by one single event i.e. the change of localization of one electron within the tubulin subunit. Gradually the conductivity of the total tubule can increase when the percentage of dimers in one specific stage increases. This opens the possibility to use the tubulin filament as a connective wire with variable conductivity

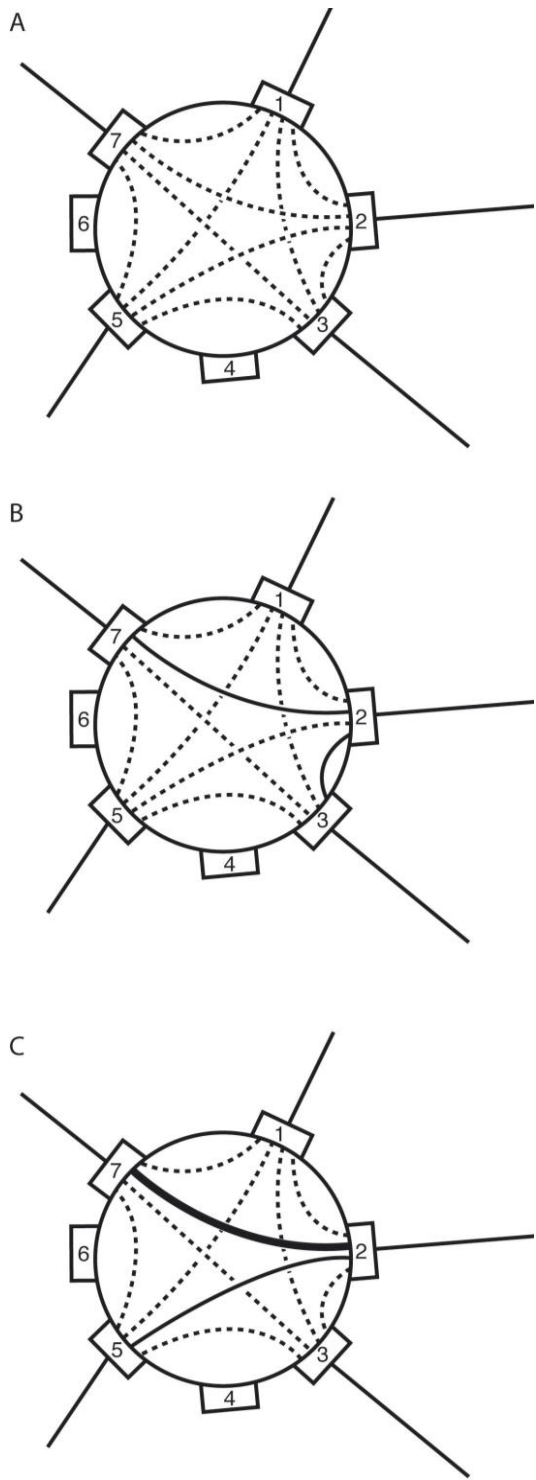

**Figure 2.**

Schematic representation of the different types of connections inside the single switchboard neuron.

**A:** All occupied cell surface determinant are potentially connected by microtubules given by dotted lines.
**B:** When two networks are associated the microtubule connection can be activated when two networks are active simultaneously. Given by the continuous line between 2 and 7.
**C:** When the association has become concrete in the process of learning the internal connection in the switchboard neuron can reach a more permanent microtubule structure. Given by the thick line between 2 and 7.

that connects different sides of the cell. That microtubules actually propagate signals was demonstrated by Vassilev et al. (1985). This results demonstrate that microtubule fibre networks may serve as an interconnecting system between membranes or membrane bounded compartments.

**Track activation**

There has to be a selection between the different activated tracks. How is a specific track selected? Here we have to think in terms of energy, as mentioned before the energy content of the tubulin dimers plays a role in the state in which it will be.
When two cell surface components are in the "on" position at the same time, there is an electrical potential between the two switch positions which will result in an activation of the tubulin connection as demonstrated by Vassilev et al. (1985). These potential differences will activate the bridge by switching all tubulin dimers in such a position that they form a connective wire. This can be imagined as a large number of compass needles that are placed in an electromagnetic field and will all take the same position resulting in a connective wire. The more units point in one direction the stronger the conductivity. The track between the two cell surface determinants is selectively activated. This is indicated in figure 2, when a dotted line changes into a solid line.
Another possibility is that of a genuine field and the compass needle metaphor becomes reality. Over the brain electromagnetic fields (John, 2001) can exist which correspond on different places and moments with different levels of energy. Becker et al. (1975) studied energy resonance transfer to different tubulin subunits or to membranes. They demonstrated that energy transfer occurs both among tubulin subunits and among these subunits and membrane proteins. The high efficiency of energy transfer indicates a considerable interaction of the tubulin and membranes. In addition, Fröhlich (1970) presents a model in which coherent excitations and cooperative coupling of proteins arrayed in an electromagnetic field may be applied to tubulin subunits.

In principle the conductivity is reproducible, it can "remember" the current which had flowed through it. Because of this quality we can consider it as a memristor (Mullin, 2009). Because of the similarity of the microtubules with the actin-myosin tubule we expect the microtubule to behave as a memristor.

All this makes it clear that so called "lost memories" can be recovered (Fisher et al. 2007). They remain intact and attached on the switchboard cell, but the internal connection is comes into the inactive stage. When recovering takes place this internal connection simply is reactivated.

**Summarizing**

We present a model that explains the storage and retrieval of memories. It is based on DNA recombination of DNA.

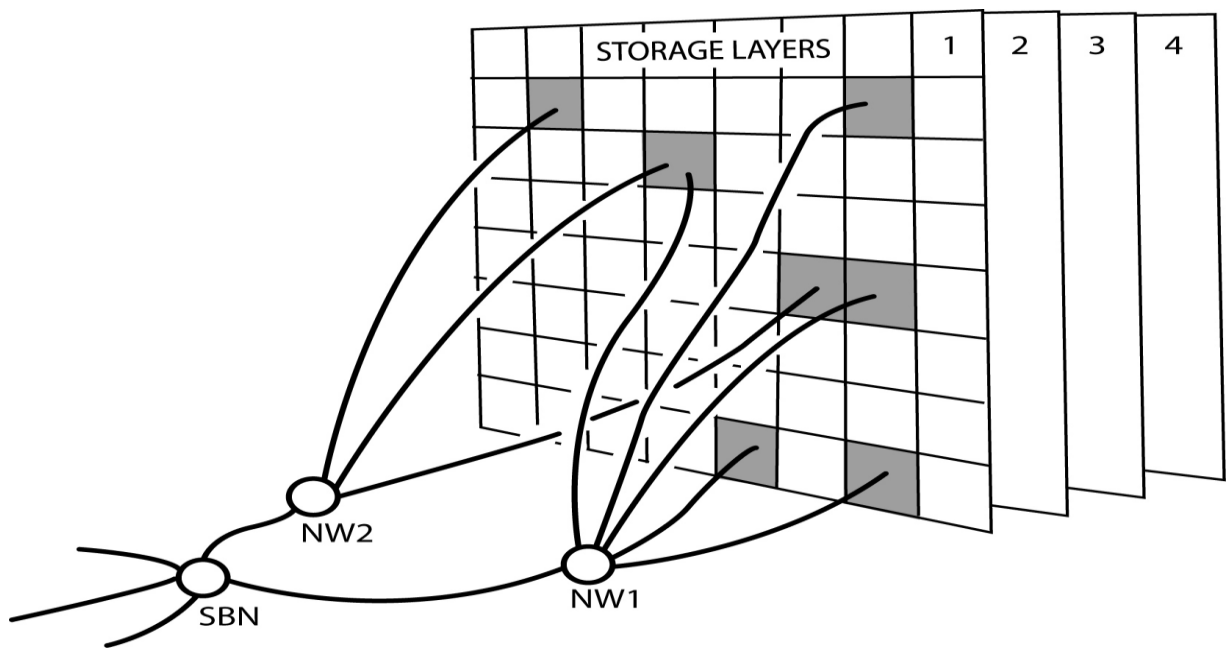

**Figure 3.**

Model of the storage of different memories in several storage layers in the cortex. Each memory is stored in a network (NW). The networks are composed of unities, which can be best imagined as pixels or voxels and one "voxel" can be used in different memories, comparable with the screen of your computer. The actual association of different memories takes place in the switchboard neuron (SBN).

In figure 3 the overall idea of our model is presented: One single memory is supposed to be composed of a number of elements. For the sake of the imagination one could see these components as the pixels or voxels of which an image is built and where each voxel can have a function in different networks. In fact there are indications that composed images, consisting of voxels are transported as a signal via a neuron to the storage location, from which it can be retrieved when necessary. The transition of a 2D image into a 1D transportable signal has been actually demonstrated by Kay et al. (2008). In figure 3 we imaged the different memories in that fashion in the storage layers which are located in the cortex. Every memory is fixed in a network (NW) in which, as far as we are concerned, a certain amount of plasticity can be tolerated. The actual association is effectuated in the switchboard neuron (SBN). Here the memristor quality of the microtubule plays a crucial role.

These association of different memories forms the mechanism which is the basis for the learning process. This way DNA recombination is used as a tool for the recombination of memories, which forms the basis for learning and eventually for consciousness.

**Acknowledgements:**

We wish to thank Mr Eelco Roos for the production of the figures.